\documentclass[12pt,manuscript]{aastex}
\usepackage{multirow}
\usepackage{lscape}
\shorttitle{The 2013 March Crab Nebula flare} 
\shortauthors{M. Mayer et al.}

\begin{document}

\title{Rapid Gamma-ray flux variability during the 2013 March Crab Nebula flare}
\author{
M.~Mayer\altaffilmark{1}, 
R.~Buehler\altaffilmark{1}, 
E.~Hays\altaffilmark{2}, 
C.~C.~Cheung\altaffilmark{3}, 
M.~S.~Dutka\altaffilmark{4}, 
J.~E.~Grove\altaffilmark{3},
M.~Kerr\altaffilmark{5}, 
R.~Ojha\altaffilmark{6,4} 
}
\altaffiltext{1}{Deutsches Elektronen Synchrotron DESY, D-15738 Zeuthen, Germany}
\altaffiltext{2}{NASA Goddard Space Flight Center, Greenbelt, MD 20771, USA}
\altaffiltext{3}{Space Science Division, Naval Research Laboratory, Washington, DC 20375-5352, USA}
\altaffiltext{4}{Catholic University of America, Washington, DC 20064, USA}
\altaffiltext{5}{W. W. Hansen Experimental Physics Laboratory, Kavli Institute for Particle Astrophysics and Cosmology, Department of Physics and SLAC National Accelerator Laboratory, Stanford University, Stanford, CA 94305, USA}
\altaffiltext{6}{ORAU/ NASA Goddard Space Flight Center, Greenbelt, MD 20771, USA}
\altaffiltext{7}{Corresponding authors: M.~Mayer, michael.mayer@desy.de.; R.~Buehler, rolf.buehler@desy.de; E.~Hays, elizabeth.a.hays@nasa.gov }

\received{2013, August 7}\accepted{2013, August 29}
\begin{abstract}
We report on a bright flare in the Crab Nebula detected by the Large Area Telescope (LAT) on board the \emph{Fermi Gamma-ray Space Telescope}. The period of significantly increased luminosity occurred in 2013 March and lasted for approximately 2 weeks. During this period, we observed flux variability on timescales of approximately $5$\,hours. The combined photon flux above 100 MeV from the pulsar and its nebula reached a peak value of $(12.5\pm 0.8)\cdot 10^{-6}$\,cm$^{-2}$\,s$^{-1}$ on 2013 March 6. This value exceeds the average flux by almost a factor of 6 and implies a $\sim20$ times higher flux for the synchrotron component of the nebula alone. This is the second brightest flare observed from this source. Spectral and temporal analysis of the LAT data collected during the outburst reveal a rapidly varying synchrotron component of the Crab Nebula while the pulsar emission remains constant in time. 
\end{abstract}


\section{Introduction}
The Crab Nebula and its pulsar are among the best-studied objects in astronomy. Their origin is assumed to be a massive core-collapse supernova observed in the year 1054 A.D. During the explosion of the progenitor star, a fast-rotating neutron star, the Crab pulsar, was formed. It emits pulsed radiation from radio wavelengths \citep[see, e.g.,][]{Hester2008} up to several hundreds of GeV \citep{Veritas2011,Magic2012}. 
The pulsar, with a spin period of 33\,ms, constantly dissipates an enormous amount of rotational energy into the surrounding medium at a rate of $4.6\cdot10^{38}$\,erg\,s$^{-1}$ \citep{Manchester2005}. A fraction of this energy  powers the acceleration of relativistic particles, which propagate away from the pulsar. These particles, thought to be mainly electrons and positrons \citep[see, e.g.,][]{Gaensler2006}, lose energy by synchrotron radiation visible from radio up to hundreds of MeV. Very high energy photons, measured up to ~50\,TeV \citep{Aharonian2006,Veritas2011,Zanin2011}, result from inverse-Compton (IC) scattering of the generated synchrotron and ambient radiation fields. 

The photon emission of the nebula powered by the relativistic particle wind is called a pulsar wind nebula (Crab Nebula in this case). Due to the underlying radiation processes a time-invariant luminosity over timescales of hundreds of years is expected. Nevertheless, instabilities in the flux of high-energy (HE; E$>100$\,MeV) $\gamma$ rays have been reported in recent years by {\it AGILE} and the {\it Fermi} Large Area Telescope ({\it Fermi}-LAT) \citep{Tavani2011,Abdo2011,Balbo2011,Striani2011,Buehler2012,Ackermann2013b}. These flares have all shown increased emission from the synchrotron component of the Crab Nebula while emission from the IC component of the nebula as well as the Crab pulsar has remained consistent with the average level.

In 2013 March, {\it Fermi}-LAT detected another bright flare of the Crab Nebula \citep{Atel}. The total flux above 100 MeV increased by almost a factor of 6, making it the second brightest flare detected from the Crab Nebula, exceeded only by the flare in 2011 April \citep{Buehler2012}. In addition to dedicated LAT observations, multi-wavelength observations were triggered to gain a more complete picture of the latest event. Detailed analyses of these observations will be published separately. In this paper, we present the temporal and spectral analyses of the {\it Fermi}-LAT data.

\section{Analysis of the Large Area Telescope data}
The {\it Fermi}-LAT is a pair-conversion detector measuring photons with energies above 20\,MeV. The point-spread function (PSF, 68\% containment radius) decreases as a strong function of energy, from $7\fdg0$  at 70\,MeV to $0\fdg25$ at 10\,GeV \citep{Ackermann2012}. Therefore, with an apparent size of $\approx 0\fdg03$, the Crab Nebula can be treated as a point source in LAT analysis. The large field of view ($\approx 2.4$\,sr) of the LAT and the survey mode used primarily for observations allow images of the full sky every three hours. 

In the beginning of 2013 March, automated science processing (ASP) of the LAT data \citep{Atwood2009, Chiang2012} detected a significantly increased photon flux of the Crab pulsar and nebula (henceforth Crab). To maximize the exposure of the Crab on short timescales, the {\it Fermi} observatory was switched to a pointed target of opportunity (ToO) observation mode between MJD 56355.65 and MJD 56359.77. For the subsequent analyses of the flare, we selected photons within a $15^{\circ}$ radius region of interest (ROI) centered on the Crab and arriving between MJD 56346.0 and MJD 56369.5 (entire analysis window, EAW). In order to avoid contamination from the $\gamma$-ray emission of the Earth limb, we only consider photons with a zenith angle less than 95$^{\circ}$. Accordingly, we did not use time windows where the edge of the ROI was located at zenith angles larger than 95$^{\circ}$. The PSF for the standard instrument response functions (IRFs) {\tt P7SOURCE\_V6} has been corrected based on flight data \citep{Ackermann2012, Ackermann2013}. The corrections effectively neglect the inclination angle dependence of the width of the PSF, which is not a consideration for analyses integrated over many orbits. However, because here we are analyzing data for short time intervals, we use the Monte-Carlo simulated {\tt P7SOURCE\_V6MC} IRFs, which include the dependence on inclination angle. We performed all analyses using unbinned {\tt gtlike} (\textit{Fermi} Science Tools v9r31p1)\footnote{Available from the {\it Fermi} Science Support Center: http://fermi.gsfc.nasa.gov/ssc/}.

We first determined the best-fit model for the EAW. The ROI source model consists of 45 sources from the 2FGL catalog (all within $20^{\circ}$ of the Crab) fixed to their published spectral parameters \citep{Nolan2012} and the models for isotropic and Galactic diffuse emission ({\tt iso\_p7v6source.txt}, {\tt gal\_2yearp7v6\_v0.fits})\footnote{\url{http://fermi.gsfc.nasa.gov/ssc/data/access/lat/BackgroundModels.html}}. In the model, we decomposed the Crab into three components, describing the pulsar as a power law with sub-exponential cut-off, the IC nebula as a smoothly-broken power law and the synchrotron component as a power law. The spectral models for the Crab with their respective parameters were adapted from \cite{Buehler2012}. Since the pulsar and IC component did not vary with time (see next paragraph), the only free parameters during the fit were the normalizations of the diffuse components, the spectral index of the power law that scales the Galactic diffuse model, and the spectral parameters of the synchrotron nebula. The spatial residuals for the fit between 70\,MeV and 300\,GeV are compatible with expectations for statistical fluctuations. The flux of the synchrotron component within the EAW was found to be $(4.05\pm 0.08)\cdot 10^{-6}\,\mathrm{cm}^{-2}\,\mathrm{s}^{-1}$, with a photon spectral index $\Gamma=3.09\pm 0.03$. Adding the constant components for the pulsar and IC nebula leads to a total Crab flux of $(6.21\pm 0.08)\cdot 10^{-6}\,\mathrm{cm}^{-2}\,\mathrm{s}^{-1}$. During this time interval, the detectable synchrotron spectrum extended to energies of about 700\,MeV. Systematic errors on integral fluxes are $\sim11\%$ and $\sim0.12$ on the spectral indicies \citep{Ackermann2012}. Using {\tt gtfindsrc}, the position of the flaring component during the EAW was found to be {R.A. = $83\fdg68$}, {Dec. = $21\fdg98$} (J2000) with a statistical error radius (68\% containment) of $0\fdg13$, which is positionally consistent with the Crab Nebula. In the following, for the temporal analyses in finer time bins, we fixed the parameters of the Galactic diffuse model to the fitted values for the EAW. The isotropic diffuse component, however, has been left free to account for variations in the cosmic-ray background, which depends on the spacecraft orbital location and orientation. 

We verified the hypothesis that the pulsar flux remains constant during the flare. We determined the ephemeris directly from the LAT data, spanning the full data set by extracting 1730 daily pulse times of arrival \citep[TOAs;][]{Ray2011} from MJD 54683 to 56425 with a typical uncertainty of $\sim$140\,$\mu$s.  Subsequently, we used {\tt tempo2} \citep{Hobbs2006} to fit a timing solution to these TOAs, using 30 harmonically-related sinusoids \citep{Hobbs2004} to model timing noise. The resulting timing solution has an RMS residual of 160\,$\mu$s, or $4.8\cdot 10^{-3}$ of the rotational period. After assigning the pulse phases $\phi$ to the photons using this ephemeris\footnote{{https://www-glast.stanford.edu/pub\_data/890/}}, we fitted the off-pulse synchrotron component ($0.5<\phi<0.8$) during the EAW, without the pulsar in the source model. The fit results agree well with the synchrotron flux determined above during the complete EAW. Using this fitted component as a fixed, underlying background, we analyzed the phase-averaged spectrum of the pulsar. Our fits are compatible with those of \cite{Buehler2012}. Therefore, within our measurement accuracy, the spectrum of the pulsar during the EAW did not change with respect to the all-time average. This was confirmed using an ephemeris derived from radio data from the Jodrell Bank Observatory \citep{Lyne1993}.

\subsection{Temporal analysis}

\begin{figure}[ht!]
\centering
\includegraphics[width=0.95\textwidth]{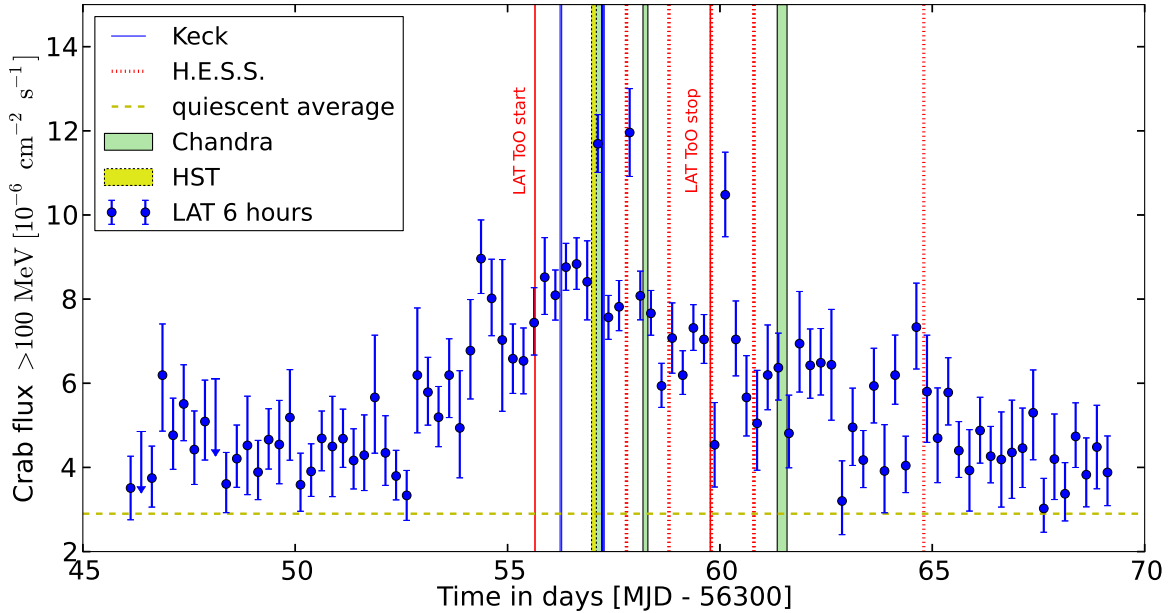}
\caption{The flux of the Crab binned in 6\,h time intervals. The fitted flux of the variable synchrotron nebula has been added to the average flux values for the Crab pulsar and IC nebula. The blue data points show the development of the Crab flux with time. Observation windows for multi-wavelength coverage provided by several instruments triggered by the {\it Fermi} measurements are overlaid in color.}
\label{fig:6hours}
\end{figure}

To obtain a time-dependent fit for the synchrotron component, we subdivided the data set into time bins of 6\,h length. The binning used here is the same as for that used in the ASP and provides a good compromise between sufficient photon statistics and sensitivity to sub-day flux variability. We analyzed the synchrotron nebula in each time interval, again leaving the pulsar and the IC nebula fixed to the values determined in \cite{Buehler2012}. 
The varying flux of the Crab for this time binning is shown in Fig.~\ref{fig:6hours}. If the test statistic (TS) for the maximum likelihood analysis \citep[see, e.g.][]{Nolan2012} was less than $4$, we calculated an upper limit (95\% confidence level, C.L.). The light curve shows three sharp spikes (MJD 56357.1, 56357.9, 56360.1) on top of a strongly increased flux level. The average synchrotron component corresponds to the all-time average determined in \cite{Buehler2012}. During the peaks, the flux was approximately twice as high as the underlying flare. Therefore, this time binning suggests an approximate flux doubling timescale of $6\,\mathrm{h}$.

\begin{figure}[t!]
\centering
\includegraphics[width=0.95\textwidth]{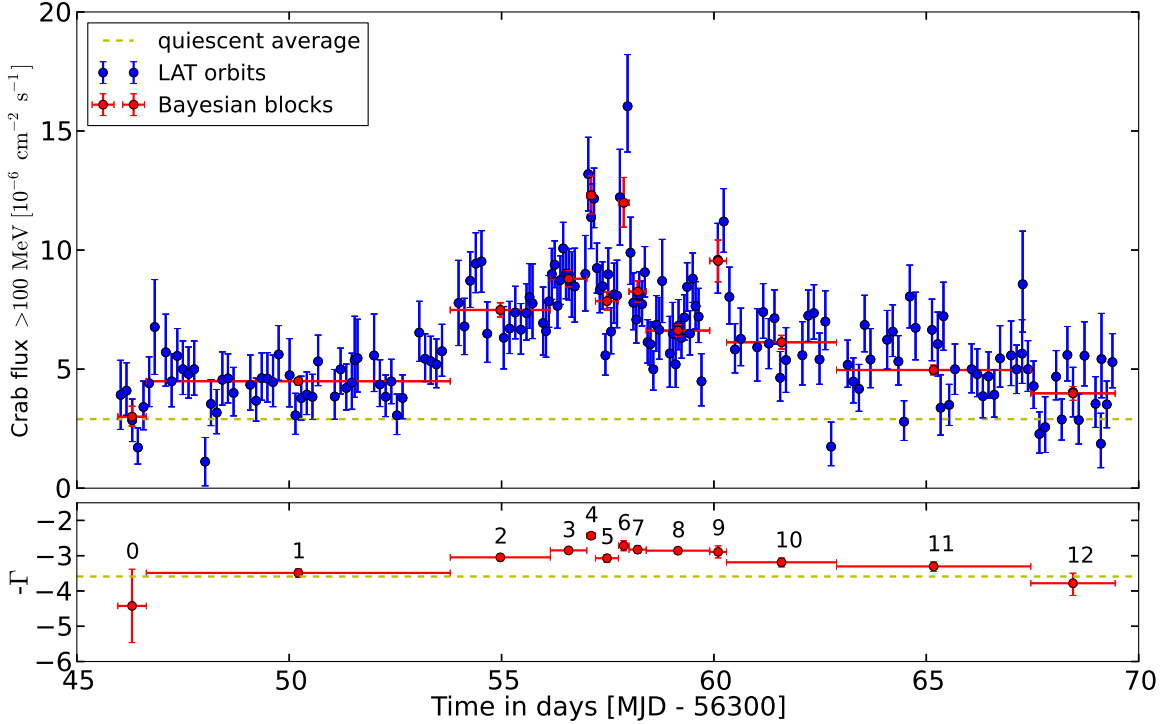}
\caption{{\it Top:} Blue data points correspond to the orbit-binned light curve while red points denote fluxes derived for the Bayesian blocks (see text). For comparison, we display the flux in the Bayesian blocks as the sum of the fitted synchrotron flux with the constant Crab pulsar and IC nebula fluxes. {\it Bottom:} Red data points show the time evolution of the spectral index of the fitted power-law model ($\frac{\mathrm{d}N}{\mathrm{d}E}\propto E^{-\Gamma}$) derived for the Bayesian blocks while the yellow-dotted line denotes the all-time average spectral index of the synchrotron component. We number the blocks here for later reference.}
\label{fig:orb}
\end{figure}

To search for more rapid flux variations, we applied a finer time binning by defining one analysis window for each orbit ($\approx 90$\,min).  In each orbit, the Crab is within the field of view of LAT for as much as $\approx 45$\,min, but that duration can be reduced significantly by details of the observing strategy from {\it Fermi}, the precession of the orbit of {\it Fermi}, and the passage of {\it Fermi} through the South Atlantic Anomaly. Many orbits thus have little or no useful exposure to the Crab. The photon statistics within these short bins do not allow a fit of the synchrotron component independently of the pulsar and inverse-Compton nebula. Therefore, we fit the spectrum of the Crab as a single power law in addition to leaving the isotropic diffuse normalization free in the fit. The light curve obtained for this binning is shown in Fig.~\ref{fig:orb}.  

To quantify the variability timescale and to assess the significance of substructures in the light curve, we decomposed it into time windows that are statistically compatible with a constant flux. These windows, so called ``Bayesian blocks'', have been obtained using the method described in \citet{Scargle1998} and are shown in Fig.~\ref{fig:orb}. For analysis in these time bins we used the same ROI and source model as for the 6\,h light curve, where the Crab is modeled as three components and the synchrotron nebula parameters are varied in the fit. The orbit-binned light curve is statistically compatible with the one obtained in the Bayesian blocks binning ($\chi^{2}/\mathrm{dof}= 180/169$), showing that substructures on shorter time scales are not statistically significant. The shortest of these time bins is 5\,h long and provides a measure of the shortest detectable variability time scale. This value is compatible with the approximate doubling time scale of 6\,h found previously. The peak flux of the Crab is $(12.5\pm 0.8)\cdot 10^{-6}$\,cm$^{-2}$\,s$^{-1}$ at the Bayesian block centered on MJD $56357.11$, almost six times larger than the average quiescent flux. If the constant flux values for the pulsar and IC nebula are subtracted, then the flux increase in the synchrotron nebula is a factor of $\sim 20$. The spectral index of a power-law model during this Bayesian block was $2.4\pm0.1$. 

Finally, we searched for variability of the synchrotron nebula on timescales shorter than the orbit binning by applying a Bayesian block analysis on the single photon arrival times. No significant short-term variability on these time scales was detected.

\subsection{Spectral analysis}

\begin{figure}[t!]
\centering
\includegraphics[width=\textwidth]{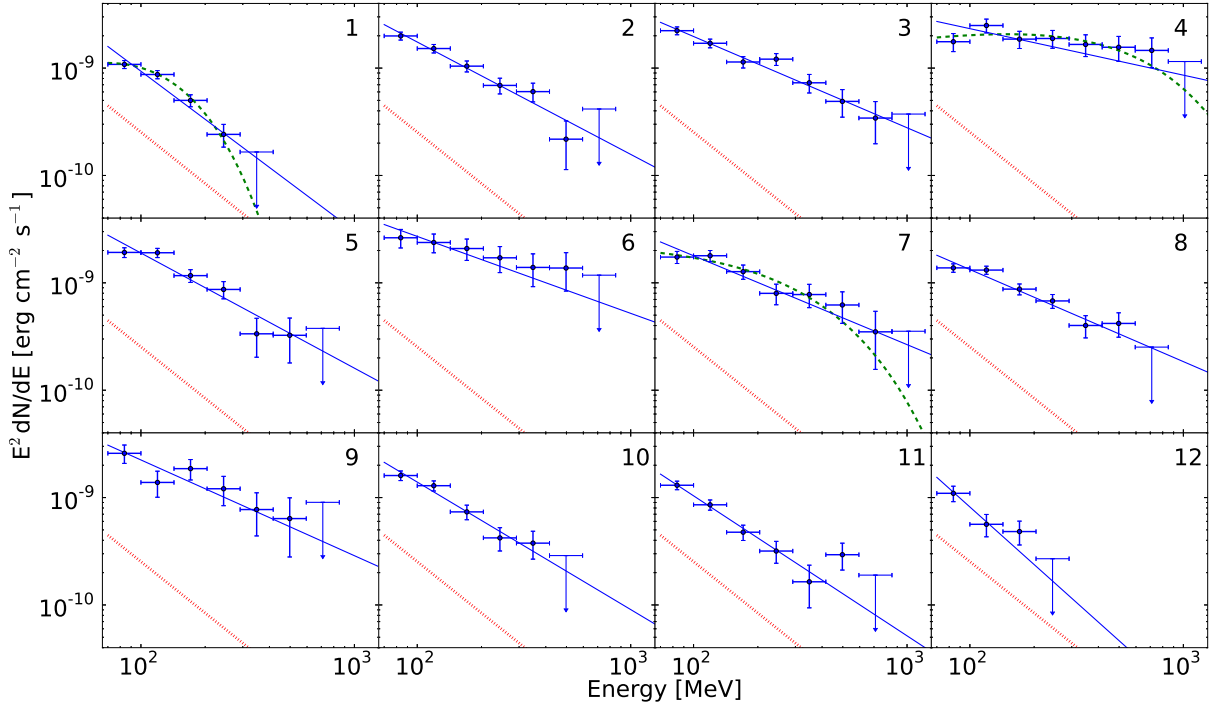}
\caption{Time-resolved spectral evolution during the flare. The energy spectra were derived in the Bayesian block binning shown in Fig. \ref{fig:orb} and described in the text. The first Bayesian block had too limited statistics for a spectral measurement and is, therefore, not shown. Data points represent flux points for the synchrotron component, while the solid lines denote the respective power-law fits. Dashed lines show the fits of a power law with an exponential cut-off if an improvement of $\Delta$TS$>9$ could be achieved by this model. Dotted lines correspond to the average synchrotron component. Upper limits have been drawn for energy bins in which the nebula is not detected, $\mathrm{TS}<4$.}
\label{fig:spectra}
\end{figure}

Similar to the 2011 April flare, the spectrum of the synchrotron nebula hardens as the flux increases. This is visualized in the bottom panel of Fig.~\ref{fig:orb}, showing the spectral indices of the power-law fits. A constant spectral index during the EAW is rejected at $7.4\sigma$. To study the spectral evolution in more detail, we determined the spectral energy distribution of the synchrotron component within each of the 13 Bayesian blocks, see Fig.~\ref{fig:spectra}. An additional component in the spectral energy distribution arises during the flaring period.
Describing the synchrotron component by power-law spectra during the flare leads to $\chi^{2}/\mathrm{dof}=61/44 $ for all data points relative to their respective model\footnote{Please note: since Bayesian block 0 has no significant data point, it was not considered in the $\chi^{2}$ calculations}. The rather poor fit probability ($\mathrm{P}=0.046$) suggests curvature in the spectra. We therefore parameterized the synchrotron nebula spectrum by a power law with an exponential cut-off, as seen during the 2011 April flare. This yields an increase $\Delta$TS$=77$ in the unbinned likelihood analysis and results in $\chi^{2}/\mathrm{dof}=37/32$. Taking into account the 13 additional degrees of freedom, this corresponds to a fit improvement on a $3.7\sigma$ level. Accordingly, we see evidence for a cut-off in the flare spectrum. Specifically, in the Bayesian blocks 1, 4 and 7 the fit quality increases with a significance $>3\sigma$. The best-fit values for these models along with the parameters of the power-law models are shown in Table~\ref{tab:parameters}. For Bayesian block 1, the cut-off is fitted below the analysis threshold energy of 70\,MeV. The results obtained in this block therefore need to be taken with care. They depend on the validity of the spectral model outside the LAT band.

In the past, {\it Fermi}-LAT has detected flares from the Crab with a variety of spectral behaviors: the flare in 2009 February exhibited only a flux increase with no spectral change. In contrast, the flares in 2010 September and 2011 April had fluxes strongly correlated with the spectral index. Moreover, in 2011 April a cut-off in the flare spectrum could be measured, which is also indicated in our flare data. In this flare, the cut-off during the brightest part of the flare (Bayesian block 4) is at $484^{+409}_{-166}$\,MeV, similar to the value of $375\pm 26$\,MeV found during the 2011 April flare \citep{Buehler2012}. 

\begin{deluxetable}{ccccc}
\tablecolumns{4}
\tablewidth{0.8\textwidth}
\tablecaption{\scriptsize{Spectral parameters of the synchrotron component for the different Bayesian blocks shown in Fig.~\ref{fig:spectra}}}
\startdata
Block number & Flux $>100$\,MeV & Photon index & Cut-off & $\Delta$TS \\
& $[10^{-6}$\,cm$^{-2}$\, s$^{-1}]$ & & $[$MeV$]$ & \\
\hline
0 & $0.8\pm 0.4$ & $4.4\pm 1.0$ & \nodata & \nodata \\ 
1 & $2.3\pm 0.2$ & $3.5\pm 0.1$ & \nodata & \nodata\\
2 & $5.3\pm 0.3$ & $3.0\pm 0.1$ & \nodata & \nodata \\ 
3 & $6.6\pm 0.3$ & $2.9\pm 0.1$ & \nodata & \nodata \\
4 & $10.2\pm 0.8$ & $2.4\pm 0.1$ &  \nodata & \nodata \\
5 & $5.7\pm 0.4$ & $3.1\pm 0.1$ & \nodata & \nodata \\
6 & $9.8\pm 1.0$ & $2.7\pm 0.1$ &  \nodata & \nodata \\
7 & $6.1\pm 0.4$ & $2.8\pm 0.1$ & \nodata & \nodata \\
8 & $4.5\pm 0.2$ & $2.9\pm 0.1$ & \nodata & \nodata \\
9 & $7.4\pm 0.9$ & $2.9\pm 0.2$ & \nodata & \nodata \\
10 & $4.0\pm 0.3$ & $3.2\pm 0.1$ & \nodata & \nodata \\
11 & $2.8\pm 0.2$ & $3.3\pm 0.1$ & \nodata & \nodata \\
12 & $1.8\pm 0.3$ & $3.8\pm 0.3$ & \nodata & \nodata \\
\hline 
1 & $2.6\pm 0.2$ & $0.7\pm 1.3$ & $53^{+28}_{-28}$ & $23.1$ \\
4 & $10.7\pm 0.8$ & $1.7\pm 0.3$ &  $484_{-166}^{+409}$ & $11.4$\\
7 & $6.5\pm 0.5$ & $2.0\pm 0.3$ &  $291_{-97}^{+192}$ & $9.2$ \\
\hline
\enddata
\tablecomments{\scriptsize{The upper section lists the results of the power-law fits, while the lower section shows the spectral parameters during the Bayesian blocks with significant cut-off ($\Delta$TS$>9$). $\Delta$TS denotes the improvement of the TS value of the cut-off model compared to the power-law fit.}}
\label{tab:parameters}
\end{deluxetable}

\section{Summary and discussion}

{\it Fermi}-LAT detected enhanced $\gamma$-ray flux from the Crab Nebula in 2013 March. The flux of the synchrotron component above 100\,MeV increased by up to a factor of $20$, making this flare the second brightest event observed to date. The shortest variability time scale was determined to be $\sim$5\,h, while the flux of the synchrotron component could be measured up to $\sim$700\,MeV. As observed in previous flares, the spectral index of the synchrotron nebula hardened significantly during periods of increased flux. An indication for an exponential cut-off in the spectrum is found at a $3.7\sigma$ level. Within the time window of the highest flux, we found a spectral index of $1.7\pm 0.3$ and a cut-off energy of $484_{-166}^{+409}$\,MeV. Interestingly, the maximum cut-off energy found in the cases of 2011 April and 2013 March is $\sim400$\,MeV.
 
Several ideas have been proposed to explain these recurring flares ($\sim$1 per year). They might originate from regions that are relativistically boosted toward us. Several works state that such regions could emerge within instabilities of the pulsar wind termination shock \citep{Bednarek2011,Komissarov2011,Lyutikov2012} or further outside at the polar region of the inner nebula \citep[e.g.][]{Lyubarsky2012}. Knots of relativistic particles could move out from the inner nebula \citep{Yuan2011} causing local, strongly varying magnetic fields \citep{Bykov2012}. 

The acceleration of the HE $\gamma$-ray emitting particles is proposed to happen via magnetic reconnection \citep[e.g.][]{Sironi2011,Uzdensky2011,Cerutti2012}. From particle-in-cell simulations, \cite{Cerutti2013} found that reconnection events can produce a flaring synchrotron component by linearly accelerating leptons within the strong electric field of the reconnection layer. This acceleration mechanism can also explain the rapid variability during the flare: beams of relativistic particles cross our line of sight and can thus cause rapid jumps in the synchrotron flux. The spectral hardening during these flux peaks can be attributed to anisotropic beaming of the relativistic particles, which causes a shift in the spectral energy distribution to higher energies. Recently, the flares have also been placed in  the context of the long-standing sigma problem of pulsar wind nebulae \citep[see, e.g.,][]{Kennel1984}. 
The flares might be part of the magnetic dissipation in the pulsar wind. As recently shown in 3D magneto-hydro dynamical simulations, magnetic dissipation can efficiently occur down stream of the wind termination shock allowing for large magnetization \citep[see, e.g.,][and references therein]{Komissarov2013,Porth2013}.

The absence of plausible counterparts at other wavelengths in past flares is one of the most surprising aspects of the flare phenomenon \citep{Weisskopf2013}. Previous non-detections of correlated variability suggested that the spectrum of the emitting electrons was hard, or possibly even mono-energetic. During the time span of the flare presented in this article, several instruments obtained simultaneous observations (see Fig.~\ref{fig:6hours}). In order to allow direct, consistent comparisons of our LAT analysis with overlapping multi-wavelength observations, we provide the corresponding spectral analysis of the LAT data for each of the observation time windows in Table~\ref{tab:table} (all spectra and light curves, along with the pulsar ephemeris applied in the analysis, are available online\footnotemark[10]{}). The results of the multi-wavelength campaign will appear in separate publications and will hopefully shed more light onto the origin of the gamma-ray flares.

\begin{landscape}
\renewcommand{\arraystretch}{0.8}
\begin{deluxetable}{c|c|cccccccc} 
\tablecolumns{10}
\tablewidth{1.13\textwidth}
\tablecaption{\scriptsize{Analysis results of LAT data corresponding to the multi-wavelength observation windows.}}
\startdata
\multirow{2}{*}{\scriptsize{Observatory}} & \scriptsize{Time} & \multicolumn{8}{c}{\scriptsize{Energy [MeV]}} \\
  & \scriptsize{[MJD-56353]} & \scriptsize{$83.7$} & \scriptsize{$119.5$} & \scriptsize{$170.7$} & \scriptsize{$243.9$} & \scriptsize{$348.5$} & \scriptsize{$497.8$} & \scriptsize{$711.1$} & \scriptsize{$1015.9$} \\
\hline
\hline
\renewcommand{\arraystretch}{0.3}
\scriptsize{HST} & \scriptsize{$3.978-4.091$} & \scriptsize{$30.4\pm 7.8$} & \scriptsize{$24.7\pm 6.9$} & \scriptsize{$14.9\pm 6.0$} & \scriptsize{$16.8\pm 6.5$} & \scriptsize{$19.2\pm 7.5$} & \scriptsize{$14.4\pm 7.8$} & \scriptsize{$22.5\pm 10.3$} & \scriptsize{$<24.9$} \\
\scriptsize{Chandra} & \scriptsize{$3.982-4.214$} & \scriptsize{$17.6\pm 3.4$} & \scriptsize{$24.9\pm 3.7$} & \scriptsize{$18.6\pm 3.4$} & \scriptsize{$18.9\pm 3.6$} & \scriptsize{$16.6\pm 3.8$} & \scriptsize{$15.6\pm 4.1$} & \scriptsize{$14.6\pm 4.5$} & \scriptsize{$<11.5$} \\
\scriptsize{Chandra} & \scriptsize{$5.188-5.304$} & \scriptsize{$17.2\pm 4.2$} & \scriptsize{$16.6\pm 4.0$} & \scriptsize{$10.3\pm 3.3$} & \scriptsize{$4.8\pm 2.9$} & \scriptsize{$5.8\pm 3.4$} & \scriptsize{$<14.5$} & \scriptsize{\nodata} & \scriptsize{\nodata} \\
\scriptsize{Chandra} & \scriptsize{$8.349-8.580$} & \scriptsize{$17.8\pm 4.8$} & \scriptsize{$7.8\pm 3.4$} & \scriptsize{$<11.0$} & \scriptsize{\nodata} & \scriptsize{\nodata} & \scriptsize{\nodata} & \scriptsize{\nodata} & \scriptsize{\nodata} \\
\scriptsize{H.E.S.S.} & \scriptsize{$4.774-4.794$} & \scriptsize{$24.4\pm 9.4$} & \scriptsize{$23.4\pm 8.8$} & \scriptsize{$25.5\pm 9.2$} & \scriptsize{$16.3\pm 8.7$} & \scriptsize{$15.8\pm 9.4$} & \scriptsize{$<27.4$} & \scriptsize{\nodata} & \scriptsize{\nodata} \\
\scriptsize{H.E.S.S.} & \scriptsize{$5.774-5.793$} & \scriptsize{$21.9\pm 9.4$} & \scriptsize{$23.4\pm 8.9$} & \scriptsize{$<24.4$} & \scriptsize{$11.0\pm 7.5$} & \scriptsize{$<22.3$} & \scriptsize{\nodata} & \scriptsize{\nodata} & \scriptsize{\nodata} \\
\scriptsize{NuSTAR} & \scriptsize{$7.319-7.507$} & \scriptsize{$11.6\pm 4.0$} & \scriptsize{$18.2\pm 4.3$} & \scriptsize{$9.7\pm 3.5$} & \scriptsize{$5.4\pm 3.0$} & \scriptsize{$6.1\pm 3.4$} & \scriptsize{$<7.6$} & \scriptsize{\nodata} & \scriptsize{\nodata} \\
\scriptsize{Keck} & \scriptsize{$3.237-3.274$} & \scriptsize{$20.8\pm 5.0$} & \scriptsize{$20.9\pm 5.0$} & \scriptsize{$14.5\pm 4.4$} & \scriptsize{$9.6\pm 4.2$} & \scriptsize{$10.1\pm 4.4$} & \scriptsize{$11.3\pm 5.3$} & \scriptsize{$<10.9$} & \scriptsize{\nodata} \\
\scriptsize{Keck} & \scriptsize{$4.238-4.275$} & \scriptsize{$17.4\pm 5.8$} & \scriptsize{$29.1\pm 6.6$} & \scriptsize{$16.0\pm 5.2$} & \scriptsize{$10.4\pm 5.0$} & \scriptsize{$<14.1$} & \scriptsize{\nodata} & \scriptsize{\nodata} & \scriptsize{\nodata} \\
\scriptsize{INTEGRAL} & \scriptsize{$0.289-5.921$} & \scriptsize{$18.9\pm 0.8$} & \scriptsize{$16.6\pm 0.7$} & \scriptsize{$11.7\pm 0.6$} & \scriptsize{$9.2\pm 0.6$} & \scriptsize{$6.3\pm 0.6$} & \scriptsize{$4.3\pm 0.6$} & \scriptsize{$2.3\pm 0.6$} & \scriptsize{$<1.4$} \\
\enddata
\tablecomments{\scriptsize{The values denote spectral flux points in fixed, logarithmic equally-spaced energy bins. The geometric mean energy of each bin is given in the first row. The second column shows the observation time window of the respective instrument. Fluxes are provided in units of $10^{-10}$\,erg\,cm$^{-2}$\,s$^{-1}$. In cases where no error is given, we list the 95\% upper limit due to $\mathrm{TS}<4$ (showing one limit after the last significant data point). Three HESS nights without simultaneous LAT coverage are not listed.}}
\label{tab:table}
\end{deluxetable}
\end{landscape}

\acknowledgments
Elizabeth Hays and Rolf B\"{u}hler acknowledge generous support from the \textit{Fermi} guest investigator program. This research was funded in part by NASA through {\it Fermi} Guest Investigator grant NNH10ZDA001N (proposal number 41213). This research was supported by an appointment to the NASA Postdoctoral Program at the Goddard Space Flight Center, administered by Oak Ridge Associated Universities through a contract with NASA. C.C.C. was supported at NRL by NASA DPR S-15633-Y. The \textit{Fermi} LAT Collaboration acknowledges generous ongoing support from a number of agencies and institutes that have supported both the development and the operation of the LAT as well as scientific data analysis. These include the National Aeronautics and Space Administration and the Department of Energy in the United States, the Commissariat \`a l'Energie Atomique and the Centre National de la Recherche Scientifique / Institut National de Physique Nucl\'eaire et de Physique des Particules in France, the Agenzia Spaziale Italiana
and the Istituto Nazionale di Fisica Nucleare in Italy, the Ministry of Education, Culture, Sports, Science and Technology (MEXT), High Energy Accelerator Research Organization (KEK) and Japan Aerospace Exploration Agency (JAXA) in Japan, and the K.~A.~Wallenberg Foundation, the Swedish Research Council and the Swedish National Space Board in Sweden.  Additional support for science analysis during the operations phase is gratefully acknowledged from the Istituto Nazionale di Astrofisica in Italy and the Centre National d’Etudes Spatiales in France.

\bibliographystyle{apj}

\end{document}